\documentclass[journal]{IEEEtran}
\usepackage[utf8]{inputenc}

\usepackage{cite}
\usepackage{amsmath,amssymb,amsfonts,amsthm}
\usepackage{graphicx}
\usepackage{algorithm}
\usepackage{algpseudocode}
\usepackage{color}
\usepackage{bm}

\newcommand{\mb}[1]{\mbox{\boldmath$#1$}}
\newcommand{\p}{\partial}
\newcommand{\beq}{\begin{eqnarray}}
\newcommand{\beqq}{\begin{eqnarray*}}
\newcommand{\eeq}{\end{eqnarray}}
\newcommand{\eeqq}{\end{eqnarray*}}

\newcommand{\x}{\mbox{\boldmath$x$}}

\newcommand{\s}{\mbox{\boldmath$s$}}

\newcommand{\w}{\mbox{\boldmath$w$}}

\newcommand{\Bb}{\mbox{\boldmath$b$}}

\newcommand{\rR}{\mathbb{R}}
\newcommand{\pP}{\mathbb{P}}
\newcommand{\eE}{\mathbb{E}}

\definecolor{red}{rgb}{1,0,0}
\DeclareMathOperator*{\argmin}{arg\,min}

\title{Numerical Twin with Two Dimensional Ornstein--Uhlenbeck Processes of Transient  Oscillations in EEG signal}
\author{P.~O.~Michel$^{1 2 3}$, C.~Sun$^{1}$, S.~Jaffard$^{2}$, D. Longrois$^{4}$ and D.~Holcman$^{13 5}$
\thanks{1 Group of Data Modeling, Computational Biology and Predictive Medicine, Institut de Biologie (IBENS), \'{E}cole Normale Sup\'{e}rieure, Universit\'{e} PSL, Paris, France.}%
\thanks{2 Univ Paris Est Creteil, Univ Gustave Eiffel, CNRS, LAMA UMR8050, F-94010 Creteil, France}%
\thanks{3 SignalMed+, 229 rue Saint-Honor\'{e}, 75001 Paris, France.}%
\thanks{4 Département d’Anesthésie, Hôpital Louis Mourier, Assistance Publique - Hôpitaux de Paris, Paris, France}%
\thanks{5 Churchill College, CB3 0DS, United Kingdom.}%
}
\date{July 2025}
\begin{document}
\maketitle

\begin{abstract}
Stochastic burst-like oscillations are common in physiological signals, yet there are few compact generative models that capture their transient structure. We propose a numerical-twin framework that represents transient narrowband activity as a two-dimensional Ornstein–Uhlenbeck (OU) process with three interpretable parameters: decay rate, mean frequency, and noise amplitude. We develop two complementary estimation strategies. The first fits the power spectral density, amplitude distribution, and autocorrelation to recover OU-parameters. The second segments burst events and performs a statistical match between empirical spindle statistics (duration, amplitude, inter-event interval) and simulated OU output via grid search, resolving parameter degeneracies by including event counts. We extend the framework to multiple frequency bands and piecewise-stationary dynamics to track slow parameter drifts. Applied to electroencephalography (EEG) recorded during general anesthesia, the method identifies OU models that reproduce $\alpha$-spindle (8–12 Hz) morphology and band-limited spectra with low residual error, enabling real-time tracking of state changes that are not apparent from band power alone. This decomposition yields a sparse, interpretable representation of transient oscillations and provides interpretable metrics for brain monitoring.
\end{abstract}


\IEEEpubid{}
\IEEEpubidadjcol

\section{Introduction} \label{sec:intro}
Noise-driven stochastic oscillations arise across physical, chemical, and biological systems. In many settings the observable signal is not a sustained sinusoid but a sequence of transient bursts that recur at irregular intervals and depart markedly from deterministic oscillators. Such behavior has been reported in seismology, biochemistry (e.g., calcium signaling dynamics \cite{Sneyd2017Dynamical}), cardiac electrophysiology \cite{Glass2020}, and neural activity, both intracellularly \cite{Gray1996Chattering,Hughes2004Synchronized,Fuentealba2005Membrane,lHorincz2009Temporal} and at the population level via local field potentials (LFP) \cite{buzsaki2006rhythms,adesnik2018cell,adamantidis2019oscillating,veit2023cortical}. In clinical electroencephalography (EEG) \cite{da1973organization}, these bursts appear as spindle-shaped events within specific frequency bands; for example, $\delta$ and $\alpha$ spindles are characteristic of sleep \cite{da1973organization,lopes1974model}, while fast ripples (100--200~Hz) have been linked to learning and memory consolidation \cite{girardeau2011hippocampal}.\\
Extracting reliable statistics from such nonstationary, narrowband burst patterns remains challenging, particularly for real-time brain monitoring during normal activity, sleep, or general anesthesia (GA) (Fig.~\ref{fig:0}). Recent CNN/RNN/transformer approaches achieve strong detection on curated sleep datasets, with reported F1 scores typically in the $0.75$–$0.90$ range and high score  for binary spindle/non-spindle decisions \cite{kulkarni2019spindlenet,muller2022waveform,mofrad2022waveform,Chen2021AutomatedSleepSpindle}. However, these methods generally deliver labels, not physiologically grounded estimators used for prediction, 
requiring a precise and robust segmentation approach \cite{Chen2021AutomatedSleepSpindle,You2024AnestheticSpindles}.\\
Our goal here is to develop a complementary, model-based representation that yields these descriptors and links them to interpretable dynamics.\\
Time–frequency and time-scale analysis (short-time Fourier and wavelet transforms \cite{Mey90I}) quantify spectral content but have limited ability to localize irregular, sporadic oscillatory structures whose envelopes vary slowly in time \cite{jaffard2001wavelets}. Empirical Mode Decomposition (EMD) can better capture modulated components \cite{flandrin2004empirical}, yet a compact, generative model that reproduces both event morphology and statistics is still lacking.\\
We shall use the narrow band two-dimensional Ornstein–Uhlenbeck (OU) process whose linear deterministic part forms a stable focus. While the process is stationary, individual realizations exhibit transient, burst-like oscillations that capture the spindle morphology observed in EEG signals. Noise-driven excursions around this focus generate transient, spindle-like activity \cite{zonca2021emergence}. The model has three interpretable parameters: decay rate $\lambda$,  a main frequency $\omega$, and noise amplitude $\sigma$, which together control event rate, coherence, and variability. We shall estimate these parameters from data using two complementary strategies:
(i) a \emph{global} fit that jointly leverages the power spectral density (PSD), marginal amplitude distribution, and autocorrelation function; and
(ii) an \emph{event-wise} fit that segments bursts (via envelope methods related to EMD \cite{huang1998empirical,flandrin2004empirical,jaffard2001wavelets,Wu2009}), then matches empirical spindle statistics (duration, amplitude, and inter-event interval) to OU simulations via grid search, with event counts resolving parameter degeneracies.
Beyond classical second-order descriptor,  we will show that spindle statistics (counts, durations, and amplitude--duration structure) impose constraints that are not captured by second-order structure alone, and are essential to resolve parameter in transient oscillatory bursts. \\
We further extend the framework to multi-band EEG by projecting the signal onto a set of coupled or independent OU processes, each describing activity in a narrow frequency band (e.g., $\alpha$ and $\delta$). A piecewise-stationary formulation enables tracking slow parameter drifts over time. As a result, the present method propose a (i) A generative ``numerical twin'' for transient oscillations based on a 2-D OU process.  (ii) Two complementary estimators—global and event-wise—that recover $(\lambda,\omega,\sigma)$.  (iii) A multi-band decomposition that represents EEG as a sparse sum of OU components plus residual background.  (iv) A real-time, sliding-window implementation that tracks slow state changes. Applied to EEG during GA, the proposed decomposition reproduces $\alpha$-spindle morphology and band-limited spectra with low residual error, and it provides measures for monitoring brain state transitions that are not apparent from band-power measures alone. We aim here at constructing a real-time numerical twin that runs in sliding windows and returns parameters for on-the-fly comparison to incoming EEG. Unlike black-box classifiers, this compact model is designed to (i) obtain trends, (ii) predict regime shifts from parameter drift, and (iii) present statistics for downstream decision support.
\section{EEG projection on two-dimensional OU-processes}
\label{sec:global_estimation}
To model burst oscillation, we use a three parameters two-dimensional Ornstein-Uhlenbeck process defined for the state variable $\s=(x,y) \in \Re^2$ as
\beq \label{eqfdt1}
\dot{\s}=A\s+ \sqrt{2\sigma}\dot{\w},
\eeq
where the anti-symmetric matrix is
\[ A=
\begin{pmatrix}
-\lambda & \omega \\
-\omega  & -\lambda \\
\end{pmatrix} \]
$\w$ is a two-dimensional Brownian motion with mean zero. { We recall that the OU-focus originates as the linear noise-driven limit of classical mass-neuronal models (Cowan–Wilson type \cite{wilson1973mathematical}) linearized around a stable operating point; random synaptic bombardment induces fluctuations that are well captured by an effective Brownian drive. Indeed, filtered shot noise generated by many independent Poisson spike trains converges to a Gaussian process with an effective variance set by the pre-synaptic rate. The Brownian noise agrees with the asymptotic limit of spiking neuronal networks \cite{wilson1973mathematical,verechtchaguina2006first,holcman2006}.
The matrix $A$ has two complex conjugated eigenvalues $\mu_{\pm}=-\lambda \pm i\omega$. Such a system generates transient fast (bursting) oscillation modulated in amplitude (Fig. \ref{fig:1}A), with a magnitude which is a few times the standard deviation $\sigma$, at a dominant resonant frequency $\omega$ (Fig. \ref{fig:1}). \\
Note that Because eq. \ref{eqfdt1} is a linear Gaussian SDE, the sampled process admits an exact discrete-time representation. Thu, an exact Gaussian likelihood for jointly estimating $(\lambda,\omega,\sigma)$  is available, but in this work segmentation is used to extract event-level spindle statistics beyond second-order structure.   we will show that event-level spindle statistics (counts, durations, and amplitude--duration structure) impose constraints that are not captured by second-order structure alone, and are essential to resolve parameter in transient oscillatory bursts. Thus our goal is to estimate the three parameters $(\lambda,\omega,\sigma)$ from empirical time series in the context of segmented spindle events: the optimal estimation constitutes the projection of the signal to the OU-process, with the optimal parameters generating an OU signal close to the EEG one, as shown in Fig.\ref{fig:0},
for an  EEG signal during general anesthesia and sleep and a  OU-process realization, which can recapitulate properties of the empirical signal.
\begin{figure}[http!]
\centering
\includegraphics[width=\linewidth]{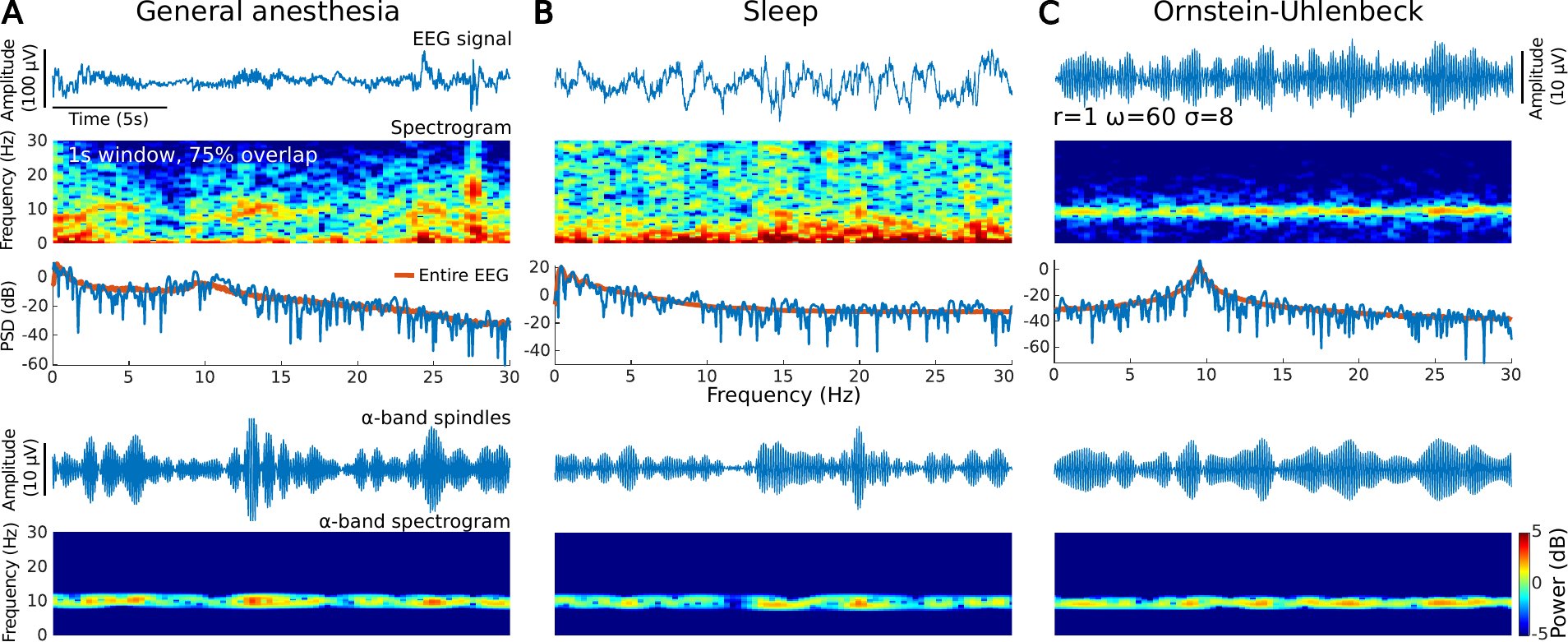}
\caption{{\bf $\alpha$ spindles dynamics during anesthesia, sleep and generated by simulations of two-dimensional Ornstein-Uhlenbeck processes.}
{\small
{\bf (A)} General Anesthesia: from top to bottom:  EEG, Spectrogram, Power Spectral Density, Filtered $\alpha$ wave, Spectrogram of the filtered EEG, showing the time-frequency spindle localization. Time-frequency representation (spectrogram) uses a 1 second window and 90\% overlap.
{\bf (B)} Same as in (A) for a sleep EEG.
{\bf (C)} Same as in (A) for a signal generated by an OU-process with parameters $r=1$, $\sigma=8$, $\omega=60$ that mimics the $\alpha$ wave signal of general anesthesia.}}
\label{fig:0}
\end{figure}
We  describe now how  the optimal parameters $\hat{\omega},\hat{\lambda},\hat{\sigma}$ of the OU process solutio of eq. \ref{eqfdt1}  are derived from  the signal $x(t)$. Briefly, we use
\begin{enumerate}
    \item Gaussian steady-state distribution of the OU-process that provide the ratio $\lambda/\sigma$
    \item Correlation function (see below) to estimate the amplitude of the noise $\sigma$.
    \item Average frequencies obtained by segmenting  burst intervals.
\end{enumerate}
{\bf \noindent Gaussian steady-state distribution to estimate the parameter ratio:} The first estimator is constructed from collecting the histogram of $x(t)$ for a sufficiently large amount of time to capture the Gaussian steady-state distribution of the process solution of eq. \ref{eqfdt1}. Indeed, the probability density function (pdf)
\beq \label{pxTtau}
p(\s,t\,|\,\s_0)\,d\x=\pP(\s(t)\in\x+d\x|\,\s_0).
\eeq
is solution of the Fokker-Planck equation (FPE) \cite{schuss2009theory}
\beq \label{FPEp}
\frac{\p p(\s,t\,|\,\s_0)}{\p t}=\sigma \Delta p(\s,t\,|\,\s_0) -\sum_{i=1}^d\frac{\p b^i(\s)p(\s,t\,|\,\s_0)}{\p x^i},
\eeq
where the vector field $\Bb(\s)=A\s$ has components
\beq
\mb{b}=
\begin{pmatrix}
-\lambda x+\omega y \\
-\omega x -\lambda y \\
\end{pmatrix}.
\eeq
While the OU Fokker–Planck operator admits an eigenfunction–eigenvalue expansion for the probability density, such spectral decompositions describe stationary transient distributions and relaxation modes. They do not capture the  inherently stochastic, short-lived bursts. Because spindles are defined by localized amplitude excursions bounded by waxing–waning envelopes, their analysis must occur in the joint time–frequency plane. The present inference relies on spectrograms and segmented event statistics rather than on global eigenfunction expansions. The steady-state \cite{risken1996fokker,schuss2009theory} of eq.\ref{FPEp} is \cite{risken1996fokker}
$ f(\s)=\frac{1}{2\pi \sqrt{\det R}}\exp\left(-\frac{1}{2} \s^T R^{-1}\s\right),$
where the matrix $R$ satisfies Lyapunov's equation $(-A)R+R (-A)^T= 2 \sigma I_{2},$
where $I_2$ is the identity matrix in dimension 2, leading to
$ R=\frac{\sigma}{\lambda} I_2$ and the steady-solution is
\beq \label{gaussianequil}
f(\s)= \frac{\lambda}{2\pi \sigma}\exp\left(-\frac{1}{2}\frac{\lambda|\s|^2}{\sigma}\right).
\eeq
To conclude, the Gaussian distribution eq. \ref{gaussianequil} allows to estimate the parameter ratio $\frac{\lambda}{\sigma}$ from  the steady-state distribution $p_{emp}(x)$ of $x(t)$, estimated over a large enough window $W_t=[t,t+T]$. The optimization procedure summarizes as
\beq
\hat{\beta} = \argmin_{\{ \beta>0 \}} \int_{x} \left(\log p_{emp}(x)-\log(\frac{1}{2\pi \beta}) +\left(\frac{x^2}{2\beta}\right)\right)^2 \mathrm{d}x,
\eeq
as illustrated in Fig.\ref{fig:1}B on a OU-process (see SI for formulation as a theorem).\\
\begin{figure}[http!]
\centering
\includegraphics[width=1\linewidth]{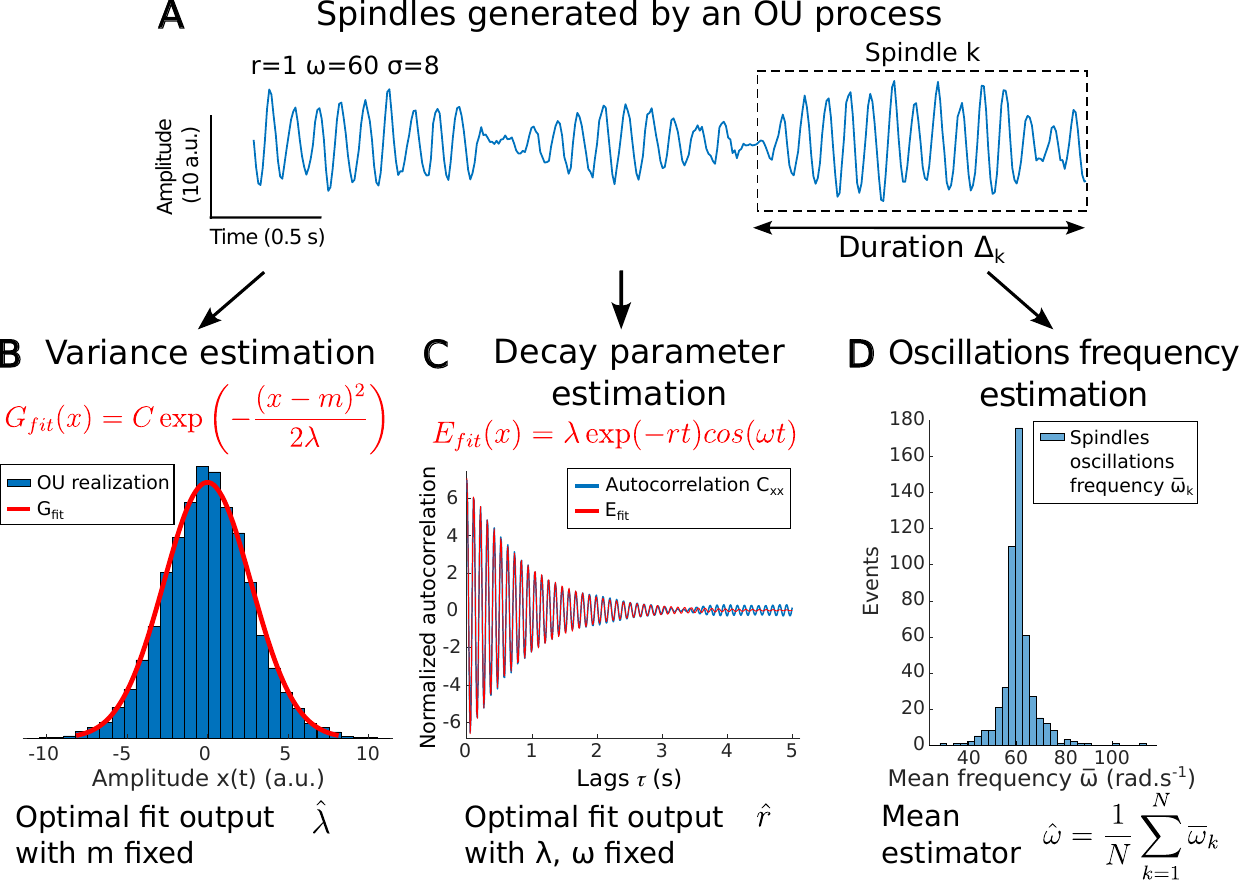}
\caption{{\bf Parameter estimation methods from a realization of a OU-process.}
{\small
{\bf (A)} Example of a signal generated by the Ornstein–Uhlenbeck (OU) model with parameters $\lambda$, $\omega$, and $\sigma$.\\
{\bf (B)} Estimator $\hat{\lambda}$ of the noise-to-decay ratio $\frac{\sigma}{\lambda}$ obtained via Gaussian fitting of the amplitude distribution.\\
{\bf (C)} Estimators of the decay coefficient $\lambda$ and noise amplitude $\sigma$ using the autocorrelation function, based on the result from (B).\\
{\bf (D)} Estimator of the frequency parameter $\omega$ using the mean instantaneous frequency computed via the Hilbert transform and averaged over all segmented spindles.
}}
\label{fig:1}
\end{figure}
At this stage, we note that, while the Ornstein--Uhlenbeck process admits well-established asymptotic theories (small-noise WKB expansions \cite{schuss2009}, Fokker--Planck eigenfunction decompositions, long-time convergence), these frameworks describe ensemble-averaged behavior and cannot directly extract time-localized transient bursts from individual sample paths. Crucially, as noise amplitude $\sigma \to 0$, the normalized power spectral density (see eq. \ref{pspectrum} below) gives $$S(f)/S(f_{\text{res}}) = \lambda^2/[\lambda^2 + (\omega - 2\pi f)^2]$$ which remains {independent} of $\sigma$: the resonance structure (peak location $\omega/(2\pi)$ and bandwidth $\lambda$) persists even as the overall amplitude vanishes, ensuring that spindles remain distinguishable from background. Eigenfunction expansions of the Fokker-Planck operator describe ensemble-averaged transient distributions and relaxation modes but miss discrete transient event structure (waxing-waning envelopes, inter-spindle intervals, counts). Moreover  small-noise asymptotics \cite{schuss2009} capture rare large deviations rather than the typical resonant fluctuations that generate spindles. We therefore develop here a hybrid approach: 1- the OU theory provides the generative model with explicit parameters $(\lambda, \omega, \sigma)$ constrained by analytic results (Gaussian steady state, autocorrelation, PSD), while the signal processing tools (envelope extraction, segmentation, spectrogram analysis) provide the empirical observables (duration, amplitude, event counts) required to identify and quantify individual bursts. This methodological synthesis is necessary because neither framework alone suffices to bridge stochastic dynamics and time-localized physiological events.
{\bf \noindent Autocorrelation function:} The second estimator uses the autocorrelation function of the component $x(t)$ of the OU-process. The power spectral density (PSD) can be computed from the Fourier transform of the autocorrelation function:
\beq \label{psd}
S(f) = \frac{1}{2\pi} \int_{\,\rR} C_{xx}(\tau) \exp(-2\pi i f \tau) \mathrm{d}\tau.
\eeq
Equation~\eqref{pspectrum} shows that the OU focus yields a pair of symmetric Lorentzian peaks centered at $\pm \omega/(2\pi)$ with width $\lambda$. In narrow bands of interest (e.g., $\delta$ and $\alpha$) the empirical PSD is dominated by such Lorentzians plus a slowly varying aperiodic background  \cite{loison2024mapping,wen_separating_2016,donoghue2020parameterizing}). Therefore, a multiscale characterization is not required for our goal: we model band-limited, quasi-monochromatic bursts with linewidth and center frequency, rather than broadband cascade dynamics. Though multiscale tools prove useful for cross-band scaling laws \cite{jaffard2001wavelets,dora2024wqn},  here we explicitly restrict to narrow band transients, where isolated Lorentzians are sufficient. Substituting the expression for the correlation function (see Appendix) associated to the OU-process
\beq \label{autocorrelation}
C_{xx}(\tau) = \frac{\sigma}{\lambda} \exp(-\lambda|\tau|) \cos(\omega \tau),
\eeq
we obtain
\begin{multline}
    S(f) = \frac{\sigma}{4\pi \lambda} \int_{\,\rR} \left( \exp(i\omega \tau) + \exp(-i\omega \tau) \right) \\ .\exp(-\lambda|\tau| - 2\pi i f\tau)\mathrm{d}\tau.
\end{multline}

Evaluating the integral leads to the explicit formula:
\beq \label{pspectrum}
S(f) = \frac{\sigma}{2\pi} \left( \frac{1}{\lambda^2 + (\omega - 2\pi f)^2} + \frac{1}{\lambda^2 + (\omega + 2\pi f)^2} \right).
\eeq
The power spectrum achieves its maximum at frequency $f_{\text{res}} = \frac{\omega}{2\pi}$. Therefore, when the parameter $\omega$ is known, the autocorrelation function can be used to extract the decay rate $\lambda$ by fitting  \eqref{autocorrelation}. In practice, after recording the one-dimensional signal $x(t)$ over a sufficiently long-time window $[t, t+T]$, subtracting the empirical mean $\langle x \rangle$, and estimating separately the dominant frequency $\omega$ from the PSD, the decay rate $\lambda$ is computed by fitting the empirical autocorrelation function of the signal: $ C_{xx}(t,s) = \eE[(x(t)-\langle x \rangle)(x(s)-\langle x \rangle)].$
This correlation is approximated by the time averaging: $\Gamma(\tau) \approx \frac{1}{T - \tau} \int_{0}^{T - \tau} x(s)\, x(s + \tau)\, \mathrm{d}s$, where $T$ must be large enough to ensure convergence to the statistical mean. This empirical autocorrelation $\Gamma(\tau)$ is fitted to \eqref{autocorrelation} to estimate $\lambda$ (Fig. \ref{fig:1}C). \\
To conclude, the autocorrelation function provides a reliable way to estimate the decay rate $\lambda$ of the OU-process; when the angular frequency $\omega_0$ and $\sigma_0$ are known, we can use a direct fit of the autocorrelation function with the empirical autocorrelation to recover the decay rate $\hat{\lambda}$ (Fig. \ref{fig:1}). This procedure is equivalent to the minimization problem
\beq \label{Optimization1}
\hat{\lambda}= \argmin_{\lambda} \int_{s\in W_t} (\Gamma(s)-R_{\lambda,\omega_0,\sigma_0}(s))^2\mathrm{d}s.
\eeq
Note that the autocorrelation function could also be used to fit the two parameters $(\lambda,\sigma)$ at the same time. 
\subsection{Spindle segmentation and estimating  the mean spindle frequency $\hat{\omega}$}
\label{s:optimization}
To estimate the mean oscillation frequency $\omega$ within each spindle, we first need to segment them. We shall use the following procedure: first, we compute the amplitude envelope of the signal $x(t)$ and then identify discrete spindle events using an empirical thresholding segmentation rule. We start with the classical chirp complex representation $x(t) = A(t) \exp(i\phi(t))$, where $A(t)$ is the instantaneous envelope that varies slowly in time, and $\phi(t)$ is the instantaneous phase capturing the fast oscillatory behavior. The envelope $A(t)$ can computed from the first step of the Empirical Mode Decomposition (EMD) \cite{huang1998empirical,flandrin2004empirical}. We define an isolated spindle pattern as an oscillatory events occurring between two consecutive local minima of the amplitude envelope, separated by a local maximum. Specifically, our segmentation procedure follows the following  steps:
\begin{enumerate}
\item \textbf{Identifying local extremum:} Detect local maxima and minima of the signal $x(t)$.
\item \textbf{Noise reduction via smoothing:} Estimate the upper $x_{\text{up}}(t)$ and lower  $x_{\text{down}}(t)$ envelopes by applying locally weighted polynomial regression (LOESS) using a second-degree model for the local maxima and minima, respectively \cite{jacoby2000loess}.
\item \textbf{Computing the instantaneous envelope distance:}
\beq
d(t) = |x_{\text{up}}(t) - x_{\text{down}}(t)|.
\eeq
\item \textbf{Thresholding extrema:} Using two thresholds $T_{\min} = \sigma_x$ and $T_{\max} = 3\sigma_x$ (where $\sigma_x$ is the empirical standard deviation of $x(t)$ previously computed),  we then collect the following sets of time points
{\small
\begin{multline}
    \mathcal{S}_{T_{\max}} = \{ t_i \,|\, \exists \varepsilon > 0,\, \forall t \in [t_i - \varepsilon, t_i + \varepsilon],\, \\ d(t_i) > d(t),\, d(t_i) > T_{\max} \},
\end{multline}
\begin{multline}
    \mathcal{S}_{T_{\min}} = \{ t_j \,|\, \exists \varepsilon > 0,\, \forall t \in [t_j - \varepsilon, t_j + \varepsilon],\, \\ d(t_j) < d(t),\, d(t_j) < T_{\min} \}.
\end{multline}

}
\item \textbf{Segmented spindles:} Segmented spindles are given by interval containing two consecutive minima $t_i, t_{i+1} \in \mathcal{S}_{T_{\min}}$, provided that the interval contains at least one local maximum $t_j \in \mathcal{S}_{T_{\max}}$:
\begin{multline} \label{spindledef}
\mathcal{S}_{\text{Spindles}} = \bigcup_{\substack{t_i, t_{i+1} \in \mathcal{S}_{T_{\min}}}}
\{ x(t)\, |\, t \in [t_i, t_{i+1}],\, \\ \exists t_j \in \mathcal{S}_{T_{\max}}, \, t_i < t_j < t_{i+1} \}.
\end{multline}
\end{enumerate}
This procedure can be  summarized as follows:
\begin{algorithm}
\caption{Real-time spindle decomposition algorithm } \label{alg:cap}
\begin{algorithmic}
\Require Acquire signal x(t) for $T=30s$
\Ensure Estimate parameters
\State decay rate $r_T$ by fitting the distribution of the signal
\State Mean frequency $\omega_T$ using Hilbert transform
\State Noise Amplitude $\sigma_T$ using autocorrelation function
\While{not found}
 \State Generate Suppression statistics, mean duration of suppression, distribution of time of successive suppressions
    \State Compare the statistics with respect to time of the EEG versus the one generate by its projection on OU process
\If{two distributions similar}
   acquire the next time window.
\ElsIf{not the same}
    \State acquire new parameters and check the trends: increase or decrease of  $(\lambda,\omega,\sigma)$.
    \State Use this to anticipate the trends of the EEG signal and Brain state.
\EndIf
\EndWhile
\end{algorithmic}
\end{algorithm}
\paragraph{Computational complexity.}
Using  the sampling frequency $F_s$ and the window duration $T$ (Algorithm 1, $T=30$\,s), so that $N=TF_s$ samples are processed per window. Using FFT-based implementations, both the Hilbert transform and autocorrelation estimation scale as $\mathcal{O}(N\log N)$ per window. The --not found-- loop (run for $I$ iterations until acceptance) performs (i) one OU simulation of length $N$ and (ii) event segmentation/statistics extraction on the window, each in $\mathcal{O}(N)$ time, followed by a distributional comparison of $S$ detected events (e.g., KS/Wasserstein after sorting) in $\mathcal{O}(S\log S)$ time with $S\leq N$. Overall, the per-window time complexity is thus
\begin{equation}
\mathcal{O}\!\left(N\log N + I\,(N + S\log S)\right)\subseteq \mathcal{O}\!\left((1+I)\,N\log N\right),
\end{equation}
and the memory is $~\mathcal{O}(N)$. Using the segmentation describe above, we extract the spindle duration, peak amplitude, and mean frequency.\\
{\noindent \bf Average spindle frequency estimator.} To compute the spindle mean frequency, we apply the Hilbert transform to the $k$-th segmented spindle $x_k(t)$
\beq
\mathcal{H}(x_k)(t) = \frac{1}{\pi} \,\text{p.v.} \int_{-\infty}^{\infty} \frac{x_k(\tau)}{t - \tau} \, \mathrm{d}\tau,
\eeq
where “p.v.” is the Cauchy principal value. From the conjugate component $y_k(t) = \mathcal{H}\{x_k\}(t)$, we have $x_{a,k}(t) = x_k(t) + i y_k(t) = A_k(t)\exp(i\phi_k(t))$, where $A_k(t)$ is the instantaneous amplitude and $\phi_k(t)$ is the instantaneous phase of spindle $k$. The instantaneous frequency is given by $ \omega_k(t) = \frac{\mathrm{d}}{\mathrm{d}t} \phi_k(t)$, which is a pointwise measure of oscillatory activity within the spindle \cite{daubechies2011synchrosqueezed}. The mean frequency of spindle $k$, segmented between time $t_k$ and $t_{k+1}$ is the average over its duration $T_k=t_{k+1}-t_{k}$:
\beq
\bar{\omega}_k = \frac{1}{T_k} \int_{t_k}^{t_{k+1}} \omega_k(t) \, \mathrm{d}t.
\eeq
The distribution of frequencies for an OU-process is shown in  Fig. \ref{fig:1}D. 
We apply the present procedure to EEG recorded from sleep and anesthesia for the three parameters $(\lambda,\omega,\sigma)$ (see also Fig. S5, where we compare the PSD of the OU-process for anesthesia and sleep using Wasserstein distance). In the case of stable  anesthesia EEG, we found $(\lambda_{An},\omega_{An},\sigma_{An})=(0.054;60;0.65)$, while for a sleep EEG, we obtain $(\lambda_{Sl},\omega_{Sl},\sigma_{Sl})=(0.045;85;0.1)$. When simulating the associated OU-processes, we obtain in the frequency band of $[7Hz,14Hz]$, the error in the power spectrum between the empirical EEG signal and optimal approximated OU-process (n=20) is given by
\beq
\frac{P_{anes}-P_{\lambda^*_{An},\omega^*_{An},\sigma^*_{An}}}{P_{anes}}&=& 0.33\pm 035\\
\frac{P_{sleep}-P_{\lambda^*_{sleep},\omega^*_{sleep},\sigma^*_{sleep}}}{P_{sleep}}&=&0.14\pm 0.29,
\eeq
see also SI Fig. S2 for a refined statistics using violin plots of the power difference and Wasserstein distance between the empirical signal and the optimal OU-realization.
\section{Projection on Multiple OU-Processes}
To extend the OU-decomposition from a single to a multiple dominant frequency band stationary signal in the linear approximation, we first identify the central frequencies for a signal $x(t)$, where the power spectral density (PSD) $P_x(f)$ exhibits isolated local maxima over a sufficiently large frequency interval.
\subsection{General projection on two-dimensional OU-processes}
For a signal that contains exactly $n_f$ dominant frequencies, the decomposition  obtained through  a  projection of $x(t)$ onto a sum of $n_f$ OU-processes, each centered at frequency $f_k$, for $k = 1, \dots, n_f$, is given by the following procedure:
\begin{enumerate}
    \item \textbf{Frequency Partitioning:} Partition the frequency axis into intervals $[a_k, a_{k+1}]$ such that within each interval the PSD $P_x(f)$ has either a unique local maximum.
    \item \textbf{Peak Detection:} For each interval $[a_k, a_{k+1}]$, identify the dominant frequency peak as
    \[
    f_k = \arg \max_{f \in [a_k, a_{k+1}]} P_x(f).
    \]
    \item On the complementary interval, $\Re^+-\bigcup_k [a_k, a_{k+1}]$, we consider that the PSD is decreasing, well approximated by a power law of the type $C_k/(b_k+f^{\alpha_k})$, where $C_k,b_k, \alpha_k$ are positive constant.
    \item \textbf{OU-Parameter Estimation:} Bandpass filter $x(t)$ in $[a_k, a_{k+1}]$ and estimate the corresponding OU-parameters $(\bar{\lambda}_k, \bar{\omega}_k, \bar{\sigma}_k)$ using the methods described in Section~\ref{s:optimization} (see also Eq.~\eqref{Optimization1}).
\end{enumerate}
The result of this procedure is a surrogate signal, sum of OU-process. The signal is the sum of each  component
\begin{equation} \label{approximationdecomposition}
x_{\mathrm{proj}}(t) \approx \sum_{k=1}^{n_f} x_k(t),
\end{equation}
where $x_k(t)$ is the OU-process with optimal parameters $(\bar{\lambda}_k, \bar{\omega}_k, \bar{\sigma}_k)$, that can estimated from the procedure described in the previous section.  Using the analytical expression for the PSD eq. \ref{pspectrum}, we obtain the total power spectrum of the projection:
\begin{align} \label{powerdecomposition}
P_{x_{\mathrm{proj}}}(f) = \sum_{k=1}^{n_f} \frac{\bar{\sigma}_k}{2\pi} \left( \frac{1}{\bar{\lambda}_k^2 + (\bar{\omega}_k - 2\pi f)^2} + \frac{1}{\bar{\lambda}_k^2 + (\bar{\omega}_k + 2\pi f)^2} \right).
\end{align}
This allow us to define the residual component
\begin{equation} \label{residual}
R(f) = P_x(f) - P_{x_{\mathrm{proj}}}(f),
\end{equation}
which captures the energy not accounted for by the OU-based projection especially outside the projection band $\rR^+-I_{proj}$. In intervals where $P_x(f)$ is monotonically decreasing, this residual will be approximated by a power-law decay:
\begin{equation} \label{decayrate}
R(f) \sim \frac{b_r}{a_r + f^r},
\end{equation}
where $a_r$ and $b_r$ are constants that depend on the local interval and $r$ is the decay exponent. In summary, the  general projection of a signal as a sum of a small number of interpretable OU-components on frequency intervals with dominant peaks, plus residual components in between represent a sparse representation for stationary signal, with three parameters for the Lorentzian decay and the decay term.
\subsection{Application to EEG projection during general anesthesia on the two $\delta-$ and $\alpha-$ frequency bands}
We apply here the OU-decomposition to the EEG signal during general anesthesia, mostly dominated by two main frequency bands: $\delta$ (0.5-4 Hz) and $\alpha$ (8-14 Hz). These bands result from random spindle patterns (Fig. \ref{fig:6}A-C).
\begin{figure}[http!]
\centering
\includegraphics[width=1\linewidth]{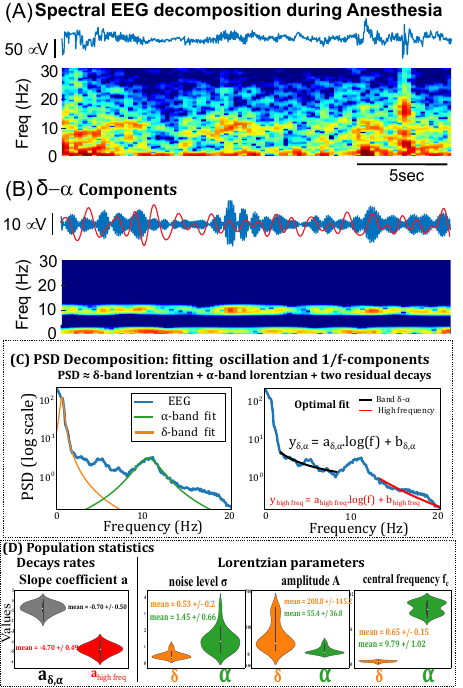}
\caption{{\bf Spectral decomposition of EEG in $\alpha-$ and $\delta-$ and residual components}
{\bf A.} EEG and spectrogram over several seconds (scal bar 5s), showing $\alpha-$ and $\delta-$ spindles.
{\bf B.}  $\alpha-$ and $\delta-$ filtered components.
{\bf C.} Power spectral decomposition (PSD).  We fitted the PSD in intermediate frequency bands $[1.8,4]$  and $[6.5,8.1]$ Hz    with $\log PSD= a \log (f)+b$.
{\bf D.} Statistical result of (a,b) coefficients for the two bands and Lorentzian parameters $\frac{A \sigma}{\pi}\frac{1}{\sigma^2+(f-f_c)^2}$, computed over n=18 EEGs.
}
\label{fig:6}
\end{figure}
For computing the approximation on the OU-processes, we directly evaluate the parameters for the $\alpha-$ (resp. $\delta-$)band in the range $[\bar{f}_{\alpha}-\Delta f_{\alpha},\bar{f}_{\alpha}+\Delta f_{\alpha}]=[8,12]$ Hz (resp. $[\bar{f}_{\delta}-\Delta f_{\delta},\bar{f}_{\alpha}+\Delta f_{\delta}]=[0,4]$ Hz) Fig. \ref{fig:6}C,  by fitting the PSD with the function
\beq
F(f)=\frac{A \sigma}{\pi}\frac{1}{\sigma^2+(f-f_c)^2}
\eeq
to the PSD for each band and obtain the distributions  shown in Fig. \ref{fig:6}D. Note that the PSD is computed on a window large enough (here ~5min). Finally, the scaling coefficient for the two decaying component is compued by  fitting the PSD in intermediate frequency bands $[1.8,4]$ (resp. $[6.5,8.1]$)Hz with $\log PSD= a \log (f)+b$, $a=-.7\pm 0.5$ (resp. $ a_{high}=-4.5\pm 0.49$). Note that this parameter estimation directly gives similar result compared to the three functions method introduced above.\\
To conclude we proposed here a decomposition of the EEG signal with OU-processes centered around a  dominant frequency band. This approximation is valid on some restricted intervals, separated by scaling law in between.
\section{Estimating the OU-parameters based on spindle statistics}
We present an alternative approach to estimate OU-parameters using the statistics
of spindle amplitude and duration segmented from the empirical time series. When the distribution of spindle durations and amplitudes are independent of the main frequency $\omega_k$ of each spindle k, which is the case for spindles found in EEG anesthesia  (Fig. S1), we focus only on identifying the two other optimal parameter $\lambda^*,\sigma^*$, assuming the mean frequency is known (for example obtained by the averaging the frequencies $\omega_k$). Increasing the sliding time window for computing spindle duration, amplitude or number leads to a converging result, as shown in Fig. S3. This is further confirmed by the duration of spindle and inter-spindle that are well approximated each by a Poisson process, as shown in Fig. S4 and S5). \\
\begin{figure}[http!]
\centering
\includegraphics[width=0.99\linewidth]{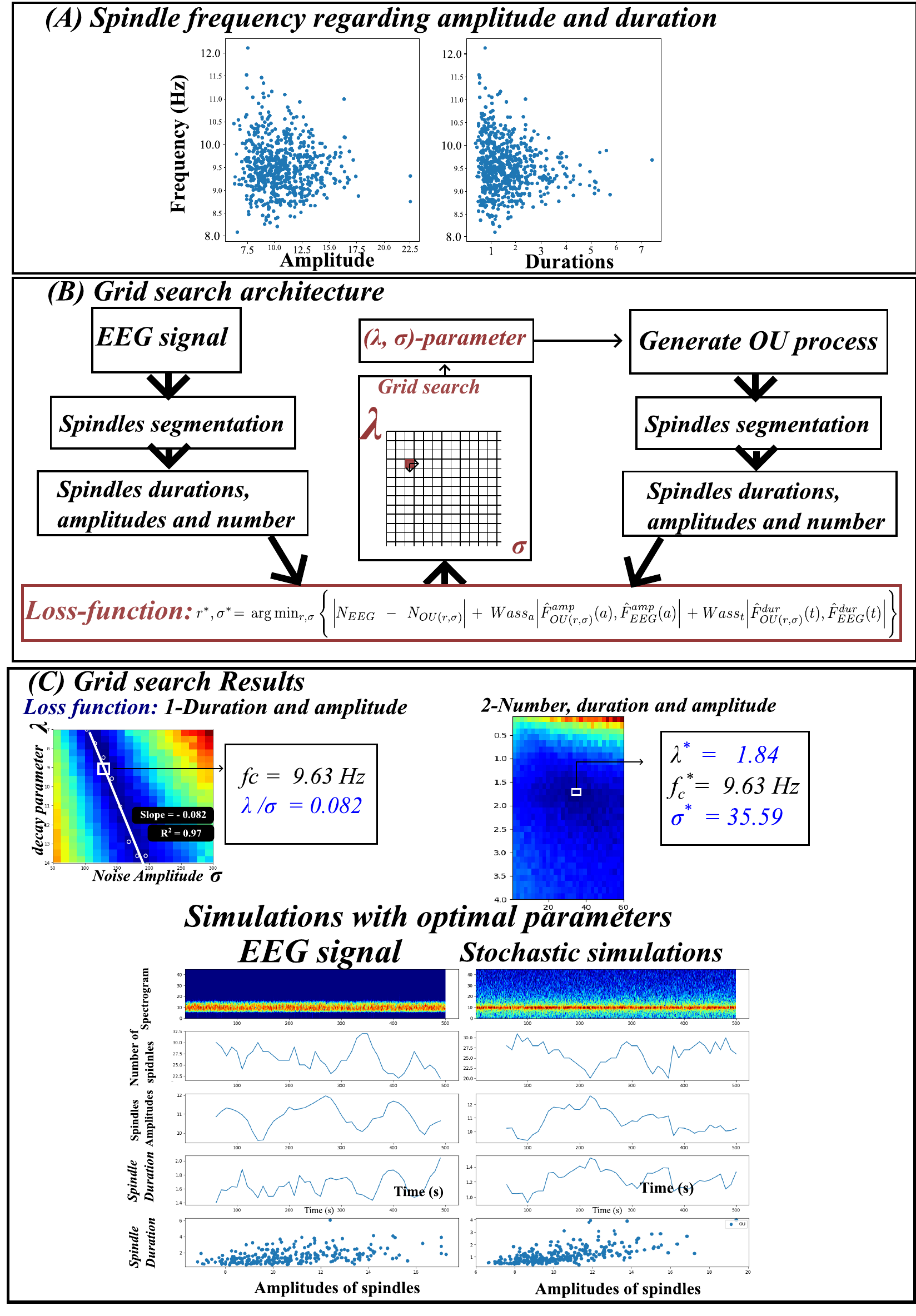}
\caption{\textbf{Parameter estimation of the stochastic OU- model to EEG spindles via grid search.}
\textbf{(A)} Parameter estimation pipeline. A grid search is performed over the parameter space of the two-dimensional Ornstein–Uhlenbeck process ($\lambda$, $\sigma$), while fixing the resonance frequency $f_c = \omega/2\pi$.
\textbf{(B)}  Grid search minimizing a loss function based on spindle statistics with (duration,  amplitude) computed from segmented EEG recordings, this leads to a line of minimum with slope $\lambda/\sigma = 0.082$. However, when considering the spindle statistics with (number, duration, and amplitude), the optimal parameters are $\lambda^* = 1.84$, $\sigma^* = 35.59$, and $f_c^* = 9.63$ Hz.
\textbf{(C)} EEG signal (top) and stochastic simulations of the OU process (bottom) using the optimal parameters.
\textbf{(D)} Spindle segmentation applied to both real EEG and simulated signals. Events are identified and classified based on amplitude and duration thresholds.
\textbf{(E)} Comparison of spindle statistics (number, amplitude, and duration) between EEG and simulations.
}
\label{fig:OU_fitting}
\end{figure}
Using a square grid search, for a fixed ensemble $(\lambda,\sigma)$ with predefined boundary, the method consists in three steps: generating OU-realizations, segmenting their spindles to obtain the duration and amplitude distribution, and comparing them with the distribution of the empirical signal (Fig. \ref{fig:OU_fitting}A).
The next point is obtained by starting again from  $(\lambda+\delta \lambda,\sigma+\delta \sigma)$.  To compare the two distribution with the empirical one, we use the Wasserstein distance \cite{friedman2001} based on the sum of absolute differences between the two cumulative density functions
\beq
Wass[\hat{F}_{OU},\hat{F}_{EEG}] = \sum_{x}(|\hat{F}_{OU}(x) - \hat{F}_{EEG}(x)|),
\eeq
where $\hat{F}(x)$ is the cumulative. Our loss function is the sum of the two Wasserstein distances for the amplitudes and durations and the absolute difference of the average number $N$ of spindles:
\beq
\begin{aligned}
\label{Wasserstein}
S_{\lambda,\sigma} = &\sum_{x}(|\hat{F}^{A}_{OU(\lambda,\sigma)}(a) - \hat{F}^{A}_{EEG}(a)|) \, + \\ & \sum_{x}(|\hat{F}^{\Delta}_{OU(\lambda,\sigma)}(t) - \hat{F}^{\Delta}_{EEG}(t)|) + |N_{OU} - N_{EEG}|
\end{aligned}
\eeq
The optimization procedure  that allows to  identify the optimal parameters is
\beq
\{\lambda^{*},\sigma^{*} \}= \argmin_{\lambda,\sigma} S_{\lambda,\sigma}.
\eeq
The results of such a procedure (Fig. \ref{fig:OU_fitting}B) depends on the parameter to be considered:  when considering the duration and amplitude only in eq. \ref{Wasserstein}, we do not find any single optimal parameter, due to a degeneracy (duration is proportional to amplitude as shown in Fig. \ref{fig:OU_fitting}B-Last raw. In that case, we obtain a line of minima where the ratio ${\sigma}/{\lambda}$ is constant (fig. \ref{fig:OU_fitting}B). However, when adding the number of spindles to the two other parameters, we remove the degeneracy and obtain a global minimum for the loss function $S_{\lambda,\sigma}$, associated with the optimal parameters (fig. \ref{fig:OU_fitting}B). Finally, we obtain a similar trend between the amplitude and duration distribution of the EEG signal vs. the optimal reconstructed OU-processes. To conclude such second approach allow to estimate OU-parameters based on the statistics of spindles.
\section{Non-stationary OU-process}
To account for a non-stationary signal such as slowly varying EEG, we introduce a time-dependent OU-process eq. \ref{eqfdt1} for the state variable $\s=(x,y) \in \rR^2$ given by
\beq
\dot{\s}=A(t)\s+ \sqrt{2\sigma(t)}\dot{\w},
\eeq
where the time-dependent anti-symmetric matrix is given by
\[ A(t)=
\begin{pmatrix}
-\lambda(t)& \omega(t) \\
-\omega(t) &-\lambda(t) \\
\end{pmatrix} \]
and $\w$ is the two-dimensional Brownian motion with mean zero. We consider the case where the parameters $(\lambda(t),\omega(t),\sigma(t))$ are slowly varying in time, so that over a time scale T, the three time-dependent parameters have their first derivative  bounded from above : $|f'(s)|,|f''(s)|$ is small for all $s\in[t,t+T]$ for $f\in\{\lambda,\omega,\sigma\}$. In practice, these functions will be piecewise constant on intervals of size $T$ so that
\beq
s\in [t,t+T] \, ,(\lambda(s),\omega(s),\sigma(s))=(\lambda_t,\omega_t,\sigma_t).
\eeq
To implement the present approach,  we estimated the three parameters on a sliding window of 60s with $\Delta T=10s$ overlapping (Fig. \ref{fig:global_ou_fit}A). Applying the estimation procedure of section \ref{sec:global_estimation} on each window, we extracted the parameters as shown in Fig. \ref{fig:global_ou_fit}B. We further identify stationary vs non-stationary time segment. Non-stationary epochs are segmented by using a sliding window of $T_{nStat}=100s$, where we computed the standard deviation $\sigma_T(t)=\int_0^t(x(s)-\bar{x})^2ds$ and the mean difference
\beq
mask(i) = \begin{cases} 1 \text{, if } sd_i > T_{std} \text{ and }  |m_{i+1} - m_i| > T_m \\ 0 \text{, otherwise} \end{cases}
\eeq
where $m_i$ is the mean of the signal on segment i. When $\sigma_T(i)>T_{std}$ and $mask(i)>T_m$, we label such segment $i$ as non-stationary. We found several stationary segment, where we  estimated the parameters of the OU-process (fig. \ref{fig:global_ou_fit}B). To conclude, the present decomposition on a OU-process can be computed for non-stationnary processes, under the condition that there are stationary epochs that are long enough.
\begin{figure}[http!]
\centering
\includegraphics[width=0.9\linewidth]{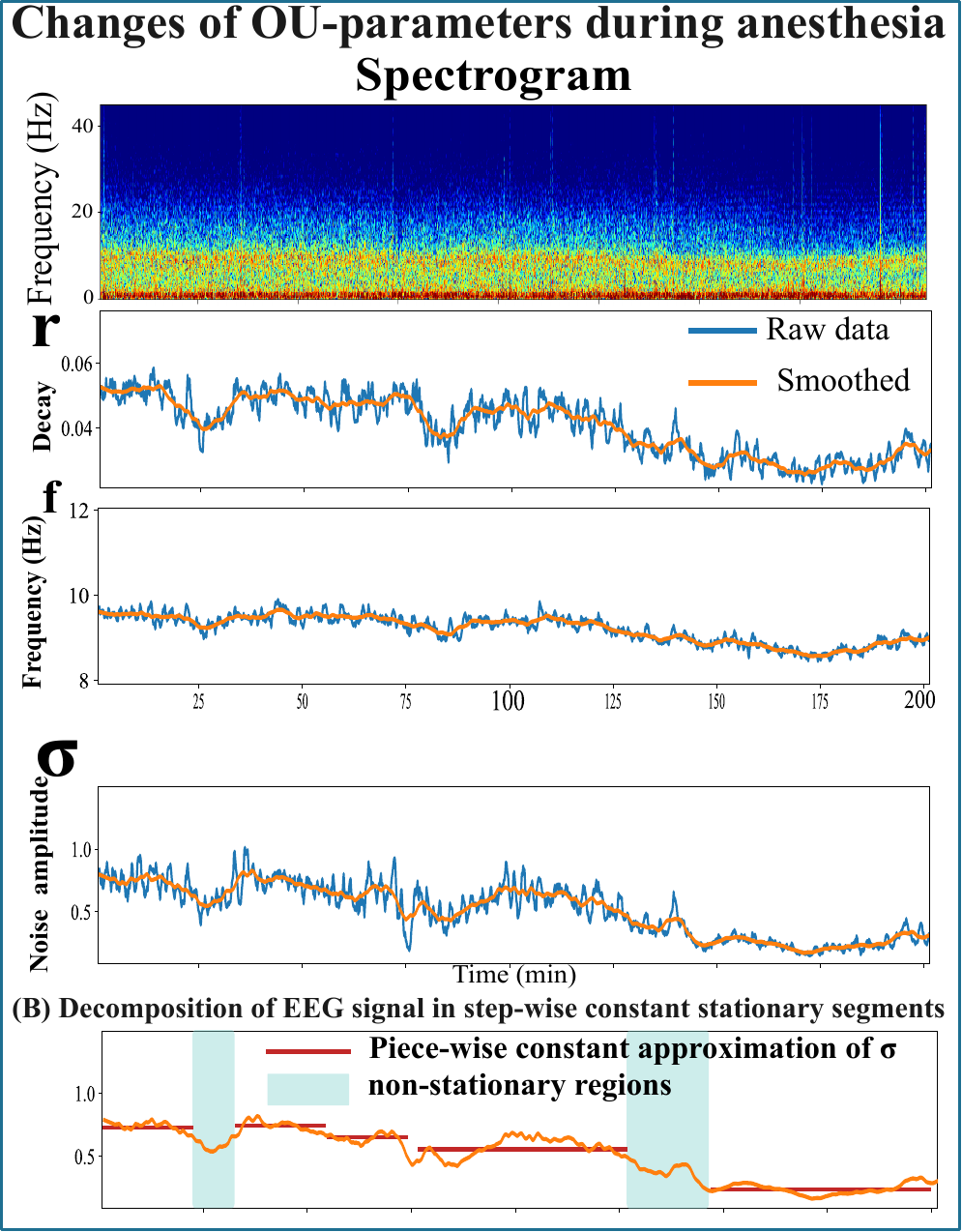}
\caption{\textbf{Non-stationary Ornstein–Uhlenbeck model.}
\textbf{(A)} Dynamics parameters $\lambda$ (decay rate), $\omega$ (rotation frequency), and $\sigma$ (noise amplitude) of the two-dimensional Ornstein–Uhlenbeck (OU) model over time. The parameters are estimated sequentially from EEG recordings filtered in the $\alpha$-band.
\textbf{(B)} Piecewise approximation of the parameter trajectories using constant segments, separated by detected transition points (black vertical lines). This segmentation captures long-range trends and identifies parameter regimes with relatively stable dynamical behavior.
\textbf{(C)} Time-frequency representation (spectrogram) of the EEG signal, showing spindle activity evolving over the course of the recording.
\textbf{(D)} Raw signal (gray) and smoothed EEG (blue) used for parameter extraction.
Together, these panels illustrate the feasibility of modeling EEG dynamics with a time-varying stochastic focus model (or equivalently linear-Gaussian continuous-time oscillator), providing a numerical twin of the patient’s brain state during anesthesia or sleep.
}
\label{fig:global_ou_fit}
\end{figure}
We can now generalize the non-stationary representation to the case of multiple dominant frequency bands so that we decompose the EEG signal as a sum
\beq \label{approximationdecomposition2}
EEG_{\hbox{reconstructed}}(t) \sim \sum_{k \in (\alpha, \delta)} \text{ou}_{k}(t) + \text{res}(t),
\eeq
where $\text{ou}_k$ are solutions of Eq. \ref{eqfdt1} with parameters $\bar{\lambda_k}, \bar\omega_k, \bar\sigma_k$  fitted to the data and $\text{res}(t)$ is a residual component. This is equivalent to the power spectral decomposition identity in a sliding window [t,t+T]
\begin{multline} \label{powerdecomposition2}
P_{EEG}(f,t) = \sum_{k \in (\alpha, \delta)} \frac{\bar\sigma_k(t)}{2 \pi} \left(\frac{1}{\bar \lambda_k^2(t)+(\bar\omega_k(t)-2\pi f)^2} \right. \\ + \left. \frac{1}{\bar \lambda_k^2(t)+(\bar\omega_k(t)+2\pi f)^2} \right) + \text{Res}(f,t),
\end{multline}
where the three parameters $(\lambda(t),\omega(t),\sigma(t))$ are piecewise constant and are estimated for each band computed on sliding windows of few minutes to obtain a local steady-state estimation, when the statistics do not vary much in time (Fig. \ref{fig:global_ou_fit}).  The residual term $\text{Res}(f,t)$  accounts for the general trend of power decay in the EEG
\beq
\text{Res}(f,t)= \frac{b_\alpha (t)}{c_\alpha(t)+f^e_{\alpha(t)}}1_{f\in B_\alpha}+\frac{b_\delta (t)}{c_\delta(t)+f^e_{\delta(t)}}1_{f\in B_\delta},
\eeq
where the frequency bands are $B_\delta=[0-4]Hz$ and $B_\alpha=[8-12]Hz$ and the coefficients $(b_\alpha,c_\alpha,e_\alpha)$ and $(b_\delta,c_\delta,e_\delta)$ are fitted to the spectrogram over a sliding window. The changes in time of the three parameters are shown in Fig. \ref{fig:global_ou_fit}.
\section{Discussion}
This work introduces a compact, generative representation for transient, narrowband oscillations based on a two-dimensional Ornstein–Uhlenbeck (OU) process with a stable focus. In contrast to descriptive time–frequency tools, the OU numerical twin provides parameterized dynamics that jointly explain morphology (waxing/waning envelopes), statistics (duration, amplitude, inter-event intervals), and spectra (Lorentzian peaks) of burst-like activity such as EEG spindles. Two complementary estimation strategies were developed: a \emph{global} route that leverages steady-state distribution, autocorrelation, and PSD, and an \emph{event-wise} route that matches segmented-burst statistics to OU simulations. Together they enable both interpretable modeling and robust tracking of slow state changes.
\subsubsection*{Relation to classical time–frequency and envelope methods}
Short-time Fourier and wavelet transforms~\cite{Mey90I} offer excellent spectral localization but do not by themselves produce a compact, mechanistic model for sporadic, irregular bursts with slow envelope modulation. Empirical Mode Decomposition (EMD) and related envelope-based approaches~\cite{flandrin2004empirical} capture instantaneous amplitudes and phases but remain descriptive. Our framework bridges these viewpoints: it uses envelope information to segment events yet constrains the dynamics with a linear stochastic focus model that yields closed-form autocorrelation and PSD expressions, making estimation principled and interpretable. Methods that separate $1/f$-aperiodic backgrounds from peaks (e.g., IRASA and parametric spectral fits~\cite{wen_separating_2016,donoghue2020parameterizing}) are complementary; here, such backgrounds are treated as residual components outside the OU bands, while peaks are fitted by OU Lorentzians with physically meaningful parameters.
\subsubsection*{Identifiability and parameter estimation}
The three OU parameters $(\lambda,\omega,\sigma)$ have distinct roles: $\omega$ sets the resonant frequency, $\lambda$ controls burst coherence and envelope decay, and $\sigma$ sets noise-driven excursion magnitude. In practice, partial identifiability issues arise if only duration and amplitude are matched; this yields a near-degeneracy along curves with approximately constant $\sigma/\lambda$. Incorporating the \emph{event count} per window breaks this degeneracy and leads to unique optima, as shown by the grid-search results. The global estimator exploits orthogonal information: (i) the steady-state Gaussian fixes $\sigma/\lambda$; (ii) the autocorrelation refines $\lambda$ given $\omega$; and (iii) the PSD peak pins down $\omega$.
\subsubsection*{Multi-band decomposition and weak coupling}
Many neural signals exhibit multiple isolated bands. We approximate each band by an independent OU component and model the remainder by smooth power-law residuals. The independence assumption is supported by the analytical cross-correlation structure of uncoupled OU processes and the weak interaction between widely separated frequencies obtained from coupled OU perturbation analysis (Appendix). In practice, the additivity of Lorentzians in~\eqref{powerdecomposition} yields a sparse spectrum model: few parameters per band plus a low-dimensional residual trend.
\subsubsection*{Tracking non-stationarity}
Brain dynamics under sleep and general anesthesia evolve over minutes. The piecewise-stationary extension maintains model simplicity while enabling slow parameter drift tracking. Because the parameters $(\lambda,\omega,\sigma)$ are estimated in sliding windows, the approach exposes clinically relevant trends that may be invisible to band-power summaries alone (e.g., an increase in $\lambda$ with nearly constant power indicates tighter burst coherence without large amplitude changes). This suggests a route to interpretable state monitoring and early-change detection.
\subsubsection*{Robustness and computational aspects}
The global estimator reduces to histogram fitting, FFT-based PSD estimation, and autocorrelation regression. The event-wise estimator depends primarily on envelope segmentation; to mitigate threshold sensitivity, we use noise-adaptive thresholds and LOESS smoothing. In our experience, parameter trajectories are stable to moderate variations in filter bandwidths and window sizes, provided windows are long enough to estimate second-order statistics (tens of seconds for EEG). The method parallelizes naturally across channels and bands, making real-time deployment feasible.\\
The study has several limitations: First, the OU focus is linear and the pdf is Gaussian. It cannot capture strongly skewed amplitude distributions, sharp waveform asymmetries, or heavy-tailed inter-event statistics; non-Gaussian extensions (e.g., Lévy or state-dependent noise) are possible future directions to explore. Second, the independence assumption between bands may be violated during cross-frequency coupling; weakly coupled OU models (Appendix) or switching linear stochastic systems could capture such regimes. Third, segmentation can fail in low SNR or with overlapping bursts; joint inference schemes that fit envelopes and OU parameters simultaneously would reduce reliance on hard thresholds.
\subsubsection*{Implications for brain monitoring}
For anesthesia and sleep, the numerical twin offers two complementary readouts. At the \emph{event} level, it yields spindle counts, durations, amplitudes, and inter-event intervals with a physiological meaning. At the model level, it provides $(\lambda,\omega,\sigma)$ trajectories that summarize latent dynamics and can be used for downstream predictors or controllers. Because the model is generative, counterfactual tests (``what if $\lambda$ increases by 20\%?'') become possible, a key ingredient for decision support, alarm design, and eventually closed-loop control.
\begin{figure*}[http!]
\centering
\includegraphics[width=0.9\linewidth]{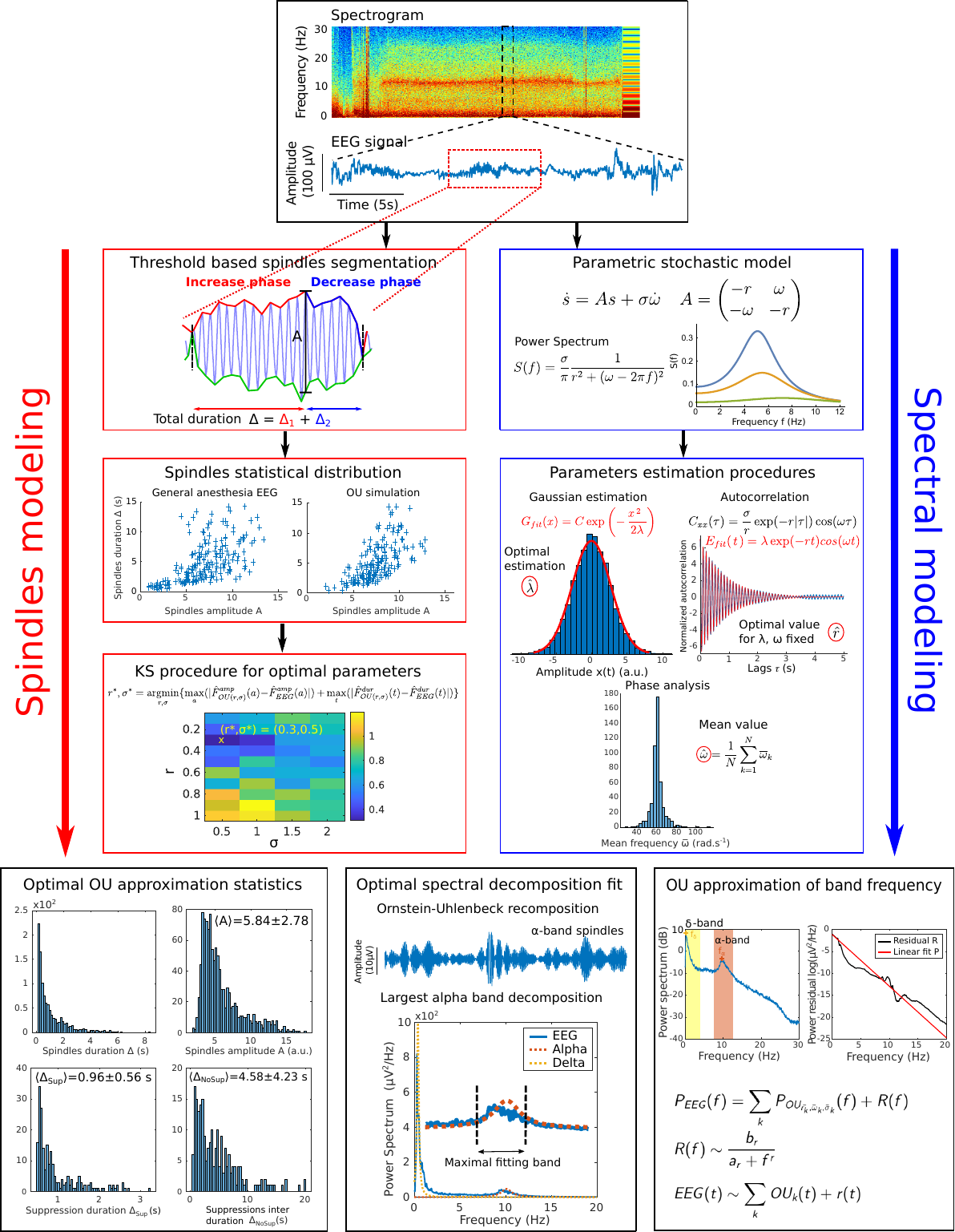}
\caption{{\bf Pipeline parameter estimation in two approaches from the EEG signal}
{\bf A.} Spectrogram of an EEG signal with magnification showing spindle events.
{\bf B.} Spindle decomposition approach. Duration and and amplitude  are extracted.
{\bf C.} Correlation distribution between spindle amplitude and duration for the EEG signal and a OU process with a given set of parameters $(\omega,\lambda,\sigma)$.
{\bf D.} Optimization procedure obtained by minimizing the sum of the Kolmogorov-Smirnov distance between the OU-process and the EEG signal for the cumulative amplitude plus duration when the two parameters $\lambda$ and $\sigma$ are free. The search of the optimum uses a greedy search in a square. {\bf E.}
{\bf F.} Alternative approach based on estimating the three model parameter from a two dimensional  OU-process with three parameters $(\omega,\lambda,\sigma)$.
{\bf G.} Estimation of the three parameters using the Gaussian distribution of the EEG signal, the correlation function and the Hilbert transform to extract the phase.
{\bf H.} Optimization allow a local optimal approximation of the power-spectrum.
{\bf I.} Decomposition of the EEG power spectrum as a sum of OU-processes plus a residual term.}
\label{fig:pip}
\end{figure*}
\section*{Acknowledgments}
We confirm that EEG were recorded from informed consent patients.
D.H. research is supported by ANR grants AstroXcite and AnalysisSpectralEEG and the European Research Council (ERC) under the European Union’s Horizon 2020 research and innovation programme (grant agreement No 882673). We use Chatgpt5.2 for correcting some sentences and verifying some computations.
\section*{Competing interests}
The Authors declare no competing financial interests.
\section*{Data availability}
The datasets generated and analysis in the current study are available from the corresponding authors upon reasonable request.
\appendices
\subsection*{Simulation of OU Processes}
The solution $x(t)$ or $y(t)$ of Eq.~(\ref{eqfdt1}) is computed  using a Euler's stochastic  scheme (Fig.~\ref{fig:0}C). For the range of considered parameters, spindles are generated.
\subsection*{Autocorrelation and PSD of the OU Focus}\label{section:autocorr}
We recall the autocorrelation function $C_{xx}(t,s)$ for the first component $x(t)$ of the Ornstein-Uhlenbeck (OU) process defined by
\beq \label{eqfdt2}
\dot{\s} = A\s + \Sigma\dot{\mb{\omega}},
\eeq
where $\mb{s} = (x,y)^T$, $A$ is the system matrix, $\Sigma = \sqrt{2\sigma} I_2$, and $\dot{\w} = (\dot{\w}_1, \dot{\w}_2)$ is a two-dimensional white Gaussian noise. The autocorrelation function of $x$ (with $\langle x(t) \rangle = 0$) is
\beq
C_{xx}(t,s) = \mathbb{E}[x(t)x(s)].
\eeq
The solution to Eq.~(\ref{eqfdt2}) is given by
\beq
\s(t) = e^{tA} \s_0 + \int_0^t e^{(t-s)A} \Sigma \, d\omega_s.
\eeq
Neglecting the initial condition in the long-time limit, we compute the covariance:
\beq
\mathbb{E}[\s(t)\s(s)^T] = 2\sigma \int_{-\infty}^{\min(t,s)} e^{A(t-t')} e^{A^T(s-t')} \, dt'.
\eeq
Using the matrix form $A = -\lambda I_2 + \omega J$, where
\[
J = \begin{pmatrix} 0 & 1 \\ -1 & 0 \end{pmatrix},
\]
we get:
\beq
\exp(At) = e^{-\lambda t}(\cos(\omega t) I_2 + \sin(\omega t) J).
\eeq
Therefore, with $\tau = t - s$:
\beq
C_{xx}(\tau) = \frac{\sigma}{\lambda} e^{-\lambda |\tau|} \cos(\omega \tau).
\eeq
Its power spectral density is obtained via Fourier transform:
\beq
\begin{aligned}
S(f) &= \frac{1}{2\pi} \int_{\mathbb{R}} C_{xx}(\tau) e^{-i2\pi f\tau} \, d\tau \\ &= \frac{\sigma}{2\pi} \left[ \frac{1}{\lambda^2 + (\omega - 2\pi f)^2} + \frac{1}{\lambda^2 + (\omega + 2\pi f)^2} \right].
\end{aligned}
\eeq
\subsection*{Weak Coupling and Cross-Correlation}
Two uncoupled bands modeled by independent OU foci are uncorrelated. With small linear coupling $\epsilon$, cross-correlation terms are $O(\epsilon)$ and decay as $e^{-\lambda|\tau|}$, with amplitude decreasing rapidly for widely separated $\omega$'s.
Consider the system of coupled Ornstein–Uhlenbeck processes:
\beq \label{coupledOU}
\Dot{X_1} &=& -\lambda X_1 + \omega_1 Y_1+ \sqrt{2\sigma_1} \Dot{w_1} + \epsilon X_2 \\
\Dot{Y_1} &=& -\omega_1 X_1 - \lambda Y_1 + \sqrt{2\sigma_1} \Dot{w_2} \nonumber\\
\Dot{X_2} &=& -\lambda X_2 + \omega_2 Y_2 + \sqrt{2\sigma_2}\Dot{w_3} \nonumber\\
\Dot{Y_2} &=& -\omega_2 X_2 -  \lambda Y_2 + \sqrt{2\sigma_2} \Dot{w_4} \nonumber
\eeq
We study the cross-correlation function between $X_1(t)$ and $X_2(s)$. The cross-correlation function is given by:
\begin{equation}
C_{13}(\tau) = \mathbb{E}[X_1(t) X_2(t-\tau)] = 2\sqrt{\sigma_1\sigma_2} \int_0^\infty \left[e^{A x} e^{A^T(x + \tau)}\right]_{1,3} dx
\end{equation}
where $A$ is the $4 \times 4$ drift matrix:
\begin{equation}
A = -\lambda I_4 + J[\omega_1, \omega_2] + \epsilon E_{13},
\end{equation}
with matrices:
\begin{align*}
J[\omega_1, \omega_2] &=
\begin{pmatrix}
0 & \omega_1 & 0 & 0 \\
-\omega_1 & 0 & 0 & 0 \\
0 & 0 & 0 & \omega_2 \\
0 & 0 & -\omega_2 & 0
\end{pmatrix}\hspace{-1mm}, &
\hspace{-3mm} E_{13} &=
\begin{pmatrix}
0 & 0 & 1 & 0 \\
0 & 0 & 0 & 0 \\
0 & 0 & 0 & 0 \\
0 & 0 & 0 & 0
\end{pmatrix}.
\end{align*}
The $(1,3)$ component of the matrix product $e^{Ax}e^{A^T(x+\tau)}$ gives the cross-correlation contribution. The cross-correlation function between \( X_1 \) and \( X_3 \) is given by:
\beq
\begin{aligned}
&C_{X_1,X_3}(\tau) = \epsilon \sqrt{\sigma_1\sigma_2} \, e^{-\lambda |\tau|}\\
&.\left(
\frac{\omega_2 \sin(\omega_2 \tau) - \omega_1 \sin(\omega_1 \tau)}{\omega_2^2 - \omega_1^2}
+ \frac{\cos(\omega_2 \tau) - \cos(\omega_1 \tau)}{\omega_2^2 - \omega_1^2}
\right)
\end{aligned}
\eeq
When $\omega_2 \gg \omega_1$,  the dominant terms in the cross-correlation function:
\begin{equation}
\begin{aligned}
C_{13}(\tau) \approx &\epsilon \sqrt{\sigma_1\sigma_2} \cdot \frac{\omega_1}{(\omega_2^2 - \omega_1^2)^2} \, \exp(-\lambda|\tau|) \\ .&\left[ \omega_2 \sin(\omega_1 |\tau|) - \omega_1 \sin(\omega_2 |\tau|) \right] + \mathcal{O}\left(\frac{1}{\omega_2^3}\right).
\end{aligned}
\end{equation}
In the limit $\omega_2 \to \infty$, this gives:
\begin{equation}
C_{13}(\tau) \sim \epsilon \sqrt{\sigma_1\sigma_2} \, \frac{\omega_1}{\omega_2^2} \, \exp(-\lambda |\tau|) \, \sin(\omega_1 |\tau|).
\end{equation}
When the frequencies are widely separated ($\omega_2 \gg \omega_1$), the cross-correlation $C_{13}(\tau)$ between the first and third components is weak and decays as $\omega_2^{-2}$. This justifies neglecting their interaction in this asymptotic regime between the $\delta$ and $\alpha$ bands and thus they could projected on two independent OU-processes.
\bibliographystyle{IEEEtran}
\bibliography{biblio}

\begin{thebibliography}{10}
\providecommand{\url}[1]{#1}
\csname url@samestyle\endcsname
\providecommand{\newblock}{\relax}
\providecommand{\bibinfo}[2]{#2}
\providecommand{\BIBentrySTDinterwordspacing}{\spaceskip=0pt\relax}
\providecommand{\BIBentryALTinterwordstretchfactor}{4}
\providecommand{\BIBentryALTinterwordspacing}{\spaceskip=\fontdimen2\font plus
\BIBentryALTinterwordstretchfactor\fontdimen3\font minus
  \fontdimen4\font\relax}
\providecommand{\BIBforeignlanguage}[2]{{%
\expandafter\ifx\csname l@#1\endcsname\relax
\typeout{** WARNING: IEEEtran.bst: No hyphenation pattern has been}%
\typeout{** loaded for the language `#1'. Using the pattern for}%
\typeout{** the default language instead.}%
\else
\language=\csname l@#1\endcsname
\fi
#2}}
\providecommand{\BIBdecl}{\relax}
\BIBdecl

\bibitem{Sneyd2017Dynamical}
J.~Sneyd, J.~M. Han, L.~Wang, J.~Chen, X.~Yang, A.~Tanimura, M.~J. Sanderson,
  V.~Kirk, and D.~I. Yule, ``On the dynamical structure of calcium
  oscillations,'' \emph{Proceedings of the National Academy of Sciences}, vol.
  114, no.~7, pp. 1456--1461, 2017.

\bibitem{Glass2020}
L.~Glass \emph{et~al.}, ``Nonlinear dynamics at the bedside: Unraveling complex
  physiological rhythms,'' \emph{Chaos: An Interdisciplinary Journal of
  Nonlinear Science}, vol.~30, no.~6, p. 061101, 2020.

\bibitem{Gray1996Chattering}
C.~M. Gray and D.~A. McCormick, ``Chattering cells: superficial pyramidal
  neurons contributing to the generation of synchronous oscillations in the
  visual cortex,'' \emph{Science}, vol. 274, no. 5284, pp. 109--113, 1996.

\bibitem{Hughes2004Synchronized}
S.~W. Hughes, M.~L{\"o}rincz, D.~W. Cope, K.~L. Blethyn, K.~A. K{\'e}kesi,
  H.~R. Parri, G.~Juh{\'a}sz, and V.~Crunelli, ``Synchronized oscillations at
  $\alpha$ and $\theta$ frequencies in the lateral geniculate nucleus,''
  \emph{Neuron}, vol.~42, no.~2, pp. 253--268, 2004.

\bibitem{Fuentealba2005Membrane}
P.~Fuentealba, I.~Timofeev, M.~Bazhenov, T.~J. Sejnowski, and M.~Steriade,
  ``Membrane bistability in thalamic reticular neurons during spindle
  oscillations,'' \emph{Journal of neurophysiology}, vol.~93, no.~1, pp.
  294--304, 2005.

\bibitem{lHorincz2009Temporal}
M.~L. L{\H{o}}rincz, K.~A. K{\'e}kesi, G.~Juh{\'a}sz, V.~Crunelli, and S.~W.
  Hughes, ``Temporal framing of thalamic relay-mode firing by phasic inhibition
  during the alpha rhythm,'' \emph{Neuron}, vol.~63, no.~5, pp. 683--696, 2009.

\bibitem{buzsaki2006rhythms}
G.~Buzsaki, \emph{Rhythms of the Brain}.\hskip 1em plus 0.5em minus 0.4em\relax
  Oxford University Press, 2006.

\bibitem{adesnik2018cell}
H.~Adesnik, ``Cell type-specific optogenetic dissection of brain rhythms,''
  \emph{Trends in Neurosciences}, vol.~41, no.~3, pp. 122--124, 2018.

\bibitem{adamantidis2019oscillating}
A.~R. Adamantidis, C.~Gutierrez~Herrera, and T.~C. Gent, ``Oscillating
  circuitries in the sleeping brain,'' \emph{Nature Reviews Neuroscience},
  vol.~20, no.~12, pp. 746--762, 2019.

\bibitem{veit2023cortical}
J.~Veit, G.~Handy, D.~P. Mossing, B.~Doiron, and H.~Adesnik, ``Cortical vip
  neurons locally control the gain but globally control the coherence of gamma
  band rhythms,'' \emph{Neuron}, vol. 111, no.~3, pp. 405--417, 2023.

\bibitem{da1973organization}
F.~L. Da~Silva, T.~Van~Lierop, C.~Schrijer, and W.~S. Van~Leeuwen,
  ``Organization of thalamic and cortical alpha rhythms: spectra and
  coherences,'' \emph{Electroencephalography and clinical neurophysiology},
  vol.~35, no.~6, pp. 627--639, 1973.

\bibitem{lopes1974model}
F.~Lopes~da Silva, A.~Hoeks, H.~Smits, and L.~Zetterberg, ``Model of brain
  rhythmic activity: the alpha-rhythm of the thalamus,'' \emph{Kybernetik},
  vol.~15, pp. 27--37, 1974.

\bibitem{girardeau2011hippocampal}
G.~Girardeau and M.~Zugaro, ``Hippocampal ripples and memory consolidation,''
  \emph{Current Opinion in Neurobiology}, vol.~21, no.~3, pp. 452--459, 2011.

\bibitem{kulkarni2019spindlenet}
R.~Kulkarni, J.~Lee, T.~R. Mullen, and G.~Cauwenberghs, ``Spindlenet: A deep
  learning model for real-time sleep spindle detection in eeg signals,''
  \emph{IEEE EMBS International Conference on Biomedical Health Informatics
  (BHI)}, pp. 1--4, 2019.

\bibitem{muller2022waveform}
L.~Müller, C.~Brunner, P.~Latuske, G.~Pipa, J.~Fell, and N.~Axmacher,
  ``Waveform detection by deep learning reveals multi-area spindles that are
  selectively modulated by memory load,'' \emph{eLife}, vol.~11, p. e75769,
  2022.

\bibitem{mofrad2022waveform}
M.~H. Mofrad, G.~Gilmore, D.~Koller, S.~M. Mirsattari, J.~G. Burneo, D.~A.
  Steven, A.~R. Khan, A.~S. Marti, and L.~Muller, ``Waveform detection by deep
  learning reveals multi-area spindles that are selectively modulated by memory
  load,'' \emph{Elife}, vol.~11, p. e75769, 2022.

\bibitem{Chen2021AutomatedSleepSpindle}
P.~Chen, D.~Chen, L.~Zhang, Y.~Tang, X.~Li \emph{et~al.}, ``Automated sleep
  spindle detection with mixed eeg features,'' \emph{Biomedical Signal
  Processing and Control}, vol.~70, p. 103026, 2021.

\bibitem{You2024AnestheticSpindles}
Y.~You, H.~Liu, Z.~Yang, Y.~Chen, F.~Yang, T.~Yu, and Y.~Zhang, ``Anesthetic
  spindles serve as eeg markers of the depth variations in anesthesia induced
  by multifarious general anesthetics in mouse experiments,'' \emph{Frontiers
  in Pharmacology}, vol.~15, p. 1474923, 2024.

\bibitem{Mey90I}
Y.~Meyer, \emph{Ondelettes et op{\'e}rateurs. I: Ondelettes}, ser. Ondelettes
  et op{\'e}rateurs.\hskip 1em plus 0.5em minus 0.4em\relax Paris: Hermann,
  1990, vol.~1.

\bibitem{jaffard2001wavelets}
S.~Jaffard, Y.~Meyer, and R.~D. Ryan, \emph{Wavelets: tools for science and
  technology}.\hskip 1em plus 0.5em minus 0.4em\relax SIAM, 2001.

\bibitem{flandrin2004empirical}
P.~Flandrin, G.~Rilling, and P.~Goncalves, ``Empirical mode decomposition as a
  filter bank,'' \emph{IEEE signal processing letters}, vol.~11, no.~2, pp.
  112--114, 2004.

\bibitem{zonca2021emergence}
L.~Zonca and D.~Holcman, ``Emergence and fragmentation of the alpha-band driven
  by neuronal network dynamics,'' \emph{PLoS Computational Biology}, vol.~17,
  no.~12, p. e1009639, 2021.

\bibitem{huang1998empirical}
N.~E. Huang, Z.~Shen, S.~R. Long, M.~C. Wu, H.~H. Shih, Q.~Zheng, N.-C. Yen,
  C.~C. Tung, and H.~H. Liu, ``The empirical mode decomposition and the hilbert
  spectrum for nonlinear and non-stationary time series analysis,''
  \emph{Proceedings of the Royal Society of London. Series A: mathematical,
  physical and engineering sciences}, vol. 454, no. 1971, pp. 903--995, 1998.

\bibitem{Wu2009}
Z.~Wu and N.~E. Huang, ``Ensemble empirical mode decomposition: A
  noise-assisted data analysis method,'' \emph{Advances in Adaptive Data
  Analysis}, vol.~1, no.~1, pp. 1--41, 2009.

\bibitem{wilson1973mathematical}
H.~R. Wilson and J.~D. Cowan, ``A mathematical theory of the functional
  dynamics of cortical and thalamic nervous tissue,'' \emph{Kybernetik},
  vol.~13, no.~2, pp. 55--80, 1973.

\bibitem{verechtchaguina2006first}
T.~Verechtchaguina, I.~M. Sokolov, and L.~Schimansky-Geier, ``First passage
  time densities in resonate-and-fire models,'' \emph{Physical Review
  E—Statistical, Nonlinear, and Soft Matter Physics}, vol.~73, no.~3, p.
  031108, 2006.

\bibitem{holcman2006}
D.~Holcman and M.~Tsodyks, ``The emergence of up and down states in cortical
  networks,'' \emph{PLoS computational biology}, vol.~2, no.~3, p. e23, 2006.

\bibitem{schuss2009theory}
Z.~Schuss, \emph{Theory and applications of stochastic processes: an analytical
  approach}.\hskip 1em plus 0.5em minus 0.4em\relax Springer Science \&
  Business Media, 2009, vol. 170.

\bibitem{risken1996fokker}
H.~Risken, ``Fokker-planck equation,'' in \emph{The Fokker-Planck
  Equation}.\hskip 1em plus 0.5em minus 0.4em\relax Springer, 1996, pp. 63--95.

\bibitem{schuss2009}
Z.~Schuss, \emph{Theory and applications of stochastic processes: an analytical
  approach}.\hskip 1em plus 0.5em minus 0.4em\relax Springer Science \&
  Business Media, 2009, vol. 170.

\bibitem{loison2024mapping}
D.~V~Loison, Necula, D.~Longrois, J.~Paz, and D.~Holcman, ``Mapping general
  anesthesia states based on electro-encephalogram transition phases,''
  \emph{NeuroImage}, vol. 285, p. 120498, 2024.

\bibitem{wen_separating_2016}
\BIBentryALTinterwordspacing
H.~Wen and Z.~Liu, ``\BIBforeignlanguage{en}{Separating {Fractal} and
  {Oscillatory} {Components} in the {Power} {Spectrum} of {Neurophysiological}
  {Signal}},'' \emph{\BIBforeignlanguage{en}{Brain Topography}}, vol.~29,
  no.~1, pp. 13--26, Jan. 2016. [Online]. Available:
  \url{http://link.springer.com/10.1007/s10548-015-0448-0}
\BIBentrySTDinterwordspacing

\bibitem{donoghue2020parameterizing}
T.~Donoghue, M.~Haller, E.~J. Peterson, P.~Varma, P.~Sebastian, R.~Gao,
  T.~Noto, A.~H. Lara, J.~D. Wallis, R.~T. Knight \emph{et~al.},
  ``Parameterizing neural power spectra into periodic and aperiodic
  components,'' \emph{Nature neuroscience}, vol.~23, no.~12, pp. 1655--1665,
  2020.

\bibitem{dora2024wqn}
M.~Dora, S.~Jaffard, and D.~Holcman, ``The wqn algorithm for eeg artifact
  removal in the absence of scale invariance,'' \emph{IEEE Transactions on
  Signal Processing}, 2024.

\bibitem{jacoby2000loess}
W.~G. Jacoby, ``Loess:: a nonparametric, graphical tool for depicting
  relationships between variables,'' \emph{Electoral studies}, vol.~19, no.~4,
  pp. 577--613, 2000.

\bibitem{daubechies2011synchrosqueezed}
I.~Daubechies, J.~Lu, and H.-T. Wu, ``Synchrosqueezed wavelet transforms: An
  empirical mode decomposition-like tool,'' \emph{Applied and computational
  harmonic analysis}, vol.~30, no.~2, pp. 243--261, 2011.

\bibitem{friedman2001}
J.~Friedman, T.~Hastie, and R.~Tibshirani, \emph{The elements of statistical
  learning}.\hskip 1em plus 0.5em minus 0.4em\relax Springer series in
  statistics New York, NY, USA:, 2001, vol.~1, no.~10.

\end{thebibliography}

\end{document}